\documentclass[trackchanges]{aastex701}

\begin{document}

    \title{Exploring the effect of mixing in Low-Luminosity Type IIp Supernovae by modeling SN~2024abfl}

\author[orcid=0009-0001-3777-0178,sname='Akshith Karri']{Akshith Karri}
\affiliation{Department of Physics and Astronomy, Purdue University, 525 Northwestern Avenue, West Lafayette, IN 47907, USA}
\email[show]{karria@purdue.edu}  

\author[orcid=0000-0002-1633-6495,sname='Abigail Polin']{Abigail Polin}
\affiliation{Department of Physics and Astronomy, Purdue University, 525 Northwestern Avenue, West Lafayette, IN 47907, USA}
\email[show]{abigail@purdue.edu}

\author[orcid=0000-0001-7626-9629,sname='Paul Duffell']{Paul Duffell}
\affiliation{Department of Physics and Astronomy, Purdue University, 525 Northwestern Avenue, West Lafayette, IN 47907, USA}
\email[show]{pduffell@purdue.edu}



\begin{abstract}

Low-luminosity Type IIp supernovae (LLSNe) are SN IIps with peak magnitudes $\geq$ -15.5 and plateau magnitudes between -13.5 and -15.5 in the V band. SN~2024abfl is an LLSN with a unique light curve, particularly the steep drop in luminosity observed after the plateau phase makes it an interesting candidate for modeling core-collapse supernova mechanisms. Using a custom pipeline involving \texttt{MESA} and \texttt{STELLA}, we investigate the possibility of suppressed ejecta mixing as a cause of the steep drop-off from the plateau phase. We find that turning off mixing mechanisms during shock breakout can mimic the distinct flat plateau and steep luminosity drop into the radioactive tail of the light curve while using previously proposed progenitor mass, radius and explosion energy parameters. Using these results as a proof-of-concept, exploring the effects of limited mixing in LLSNe candidates could give us better insight into how they differ from Typical Type IIp SNe.

\end{abstract}

\keywords{\uat{Supernovae}{1668} --- \uat{Computational Astronomy}{293} --- \uat{Core-colllapse supernovae}{304} --- \uat{Type II supernovae}{1731}}


\section{Introduction} \label{sec:introduction}

SN 2024abfl is an unusual low-luminosity Type IIp supernova exhibiting a remarkably flat $\sim126$ day plateau followed by a rapid decline of $\sim2.4$ mag in only $\sim5.5$ days onto a faint radioactive tail \citep{gerard2026sn2024abflflatplateaulowluminosity}. The extended low-luminosity plateau and dim nickel tail suggest a distinct explosion and progenitor configuration. The distinct drop in luminosity during the transition from the plateau phase to the radioactive nickel tail when compared with other well-studied LLSNe like SN 2003Z or SN 2005cs \citep{10.1093/mnras/stu156}, where this transition appears less steep, is another clue to the unique structure of this event.

Interpreting these observables is complicated by the uncertain distance to the host galaxy NGC 2146 \citep{callis2021lowgosn2018zd}. Because the galaxy is sufficiently nearby for peculiar velocities to affect redshift-based distance estimates, yet distant enough to limit the precision of Cepheid measurements, the inferred luminosity of SN 2024abfl remains poorly constrained. These uncertainties introduce significant degeneracies in the physical parameters inferred from lightcurve modeling, including the progenitor zero-age main-sequence mass, progenitor radius and explosion energy.

In this work, we model the photometric evolution of SN~2024abfl using a custom pipeline involving \texttt{MESA} and \texttt{STELLA} \citep{2026AAS...24727702K}. We focus our analysis on the unusual plateau morphology and transition to the radioactive tail in order to explore the extent to which distinct progenitor and explosion configurations, and in particular, the amount of mixing in the ejecta, can reproduce the observed properties of SN 2024abfl.

\section{Methods} \label{sec:methods}

We model the progenitor evolution and explosion using a custom pipeline involving \texttt{MESA} \& \texttt{STELLA}, which consists of three relevant steps. \texttt{MESA} \citep[][]{Paxton2019} is used to evolve one-dimensional stellar progenitor models while simultaneously solving the stellar structure and composition equations with appropriate equations of state and nuclear reaction networks. The resulting progenitor structures are exploded and evolved through shock breakout using a modified built-in version of \texttt{STELLA} \citep{2004Ap&SS.290...13B, 2005AstL...31..429B, 2006A&A...453..229B, 2019ApJ...879....3G}, which performs coupled hydrodynamic and radiative-transfer calculations to produce synthetic lightcurves.

The progenitor is first created and evolved from Pre-Main Sequence to Core Collapse using the \texttt{MESA} \texttt{20M\_pre\_ms\_to\_core\_collapse} test suite. This step allows us to vary initial parameters like initial mass, metallicity, wind schemes, and \texttt{mixing\_length\_alpha}, which is multiplied with the local pressure scale height to give the mixing length, and is inversely proportional to the radius-luminosity ratio. The core-collapse is then carried out using a modified version of the \texttt{ccsn\_IIp} test suite, and allows us to modify the explosion energy and nickel mass of the ejecta. The test suite also provides options to toggle hydrodynamics, RTI mixing, and boxcar mixing. \texttt{MESA}'s in-built version of \texttt{STELLA} finally allows us to complete shock breakout and evolve the ejecta until an epoch of our choosing, which was set to 400 days.

We compare our results with Bolometric Light Curve data from \cite{gerard2026sn2024abflflatplateaulowluminosity}, which was generated from photometric data using the Light Curve Fitting Package \citep{hosseinzadeh_2024_11405219}.

\section{Results} \label{sec:results}

\begin{figure}[ht]
    \centering
    \includegraphics[width=0.95\linewidth]{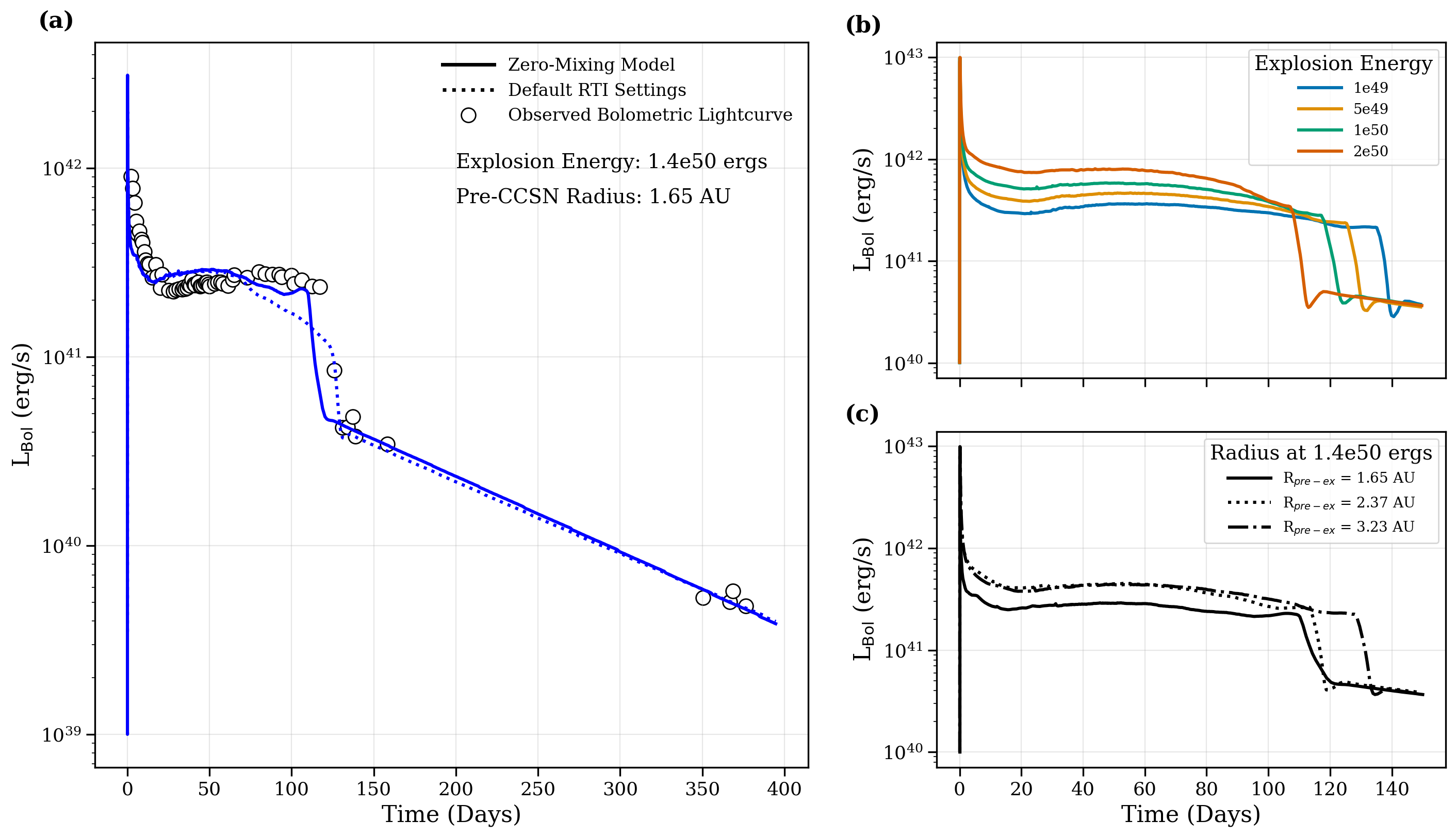}
    \caption{\textbf{(a)} Bolometric light-curve fit to SN 2024abfl (white points), showing models computed with (dashed line) and without (solid line) RTI-induced mixing. \textbf{(b)} Effects of varying explosion energy while holding other model parameters fixed, demonstrating how luminosity drops while extending the plateau phase duration as explosion energy decreases. \textbf{(c)} Effects of varying progenitor radius showing how a decrease in radius shrinks both plateau time and luminosity. Plots \textbf{(b)} and \textbf{(c)} are models without any mixing in the ejecta during the shock breakout phase.}
    \label{fig:Figure 1}
\end{figure}

Figure 1 (a) shows our final model, consisting of a 10 $M_{\odot}$ progenitor with solar metallicity, an explosion energy of $1.4\times10^{50}$ ergs, a synthesized $^{56}$Ni mass of 0.01 $M_{\odot}$, and a pre-collapse stellar radius of $1.65$ AU ($\approx 355,R_{\odot}$). This model reproduces the overall morphology of the observed Type II supernova light curve, including the plateau phase and subsequent transition to the radioactive tail, while staying in agreement with proposed progenitor conditions in \cite{gerard2026sn2024abflflatplateaulowluminosity}.

To investigate parameter sensitivities, we computed a suite of models varying the explosion energy while holding all other parameters fixed (Figure \ref{fig:Figure 1} (b)). Increasing the explosion energy produces systematically brighter light curves with shorter plateau durations. This behavior is consistent with more energetic explosions depositing additional energy into the ejecta while accelerating the recombination-driven evolution of the expanding envelope.

Figure 1 (c) illustrates the effect of varying the progenitor radius. Decreasing the progenitor radius results in both lower luminosities and shorter plateau durations while preserving the overall morphology of the light curve. The reduced radiating surface area and lower thermal energy reservoir of more compact progenitors naturally lead to dimmer and more rapidly evolving events.

\subsection{Effects of Mixing on Light Curve Evolution}

To investigate the origin of the unusually sharp plateau-to-tail transition, we explored the role of ejecta mixing in shaping the light curve. During the plateau phase, SN~2024abfl exhibits an exceptionally flat evolution in both the \textit{V} and \textit{r} bands, followed by a rapid decline of approximately $2.4$ mag over $\sim5.5$ days into the radioactive $^{56}$Ni-powered tail \citep{gerard2026sn2024abflflatplateaulowluminosity}. In contrast, most Type II supernovae display a more gradual transition between the plateau and tail. This smoother evolution is generally attributed to efficient mixing of the ejecta \citep{2017ApJ...846...37U}, which redistributes radioactive material and thermal energy throughout the expanding envelope.

The unusually flat plateau of SN~2024abfl suggests that there is little contribution from the deeper ejecta during the hydrogen recombination phase. As the hydrogen envelope completes recombination, the luminosity transitions almost directly to emission powered by the decay of $^{56}$Ni and its daughter products, producing the observed sharp luminosity drop.

Consistent with this interpretation, \texttt{MESA} models computed without any artificial boxcar smoothing or Rayleigh--Taylor instability (RTI) mixing produce a substantially steeper decline at the end of the plateau than models that include mixing. The unmixed models exhibit a rapid luminosity drop approximately $125$ days after explosion that more closely matches the observed light curve of SN~2024abfl. Although the real ejecta are unlikely to be completely unmixed, these results suggest that significantly less mixing is necessary to reproduce the morphology of the plateau-to-tail transition.

\section{Discussion} \label{sec:discussion}

Though nearby circumstellar interactions affecting the early-time light curve were not considered, our simulations demonstrate that the degree of ejecta mixing significantly affects the morphology of the plateau-to-tail transition. In the limiting case of no mixing, where both boxcar smoothing and Rayleigh--Taylor instability (RTI) mixing are disabled, the model exhibits a sharp luminosity decline at the end of the plateau phase. The observed differences between the RTI and non-RTI models suggest that a scenario with limited mixing may provide a better representation of SN 2024abfl than models with extensive mixing during the shock breakout phase.

The trends observed throughout our model grid are broadly consistent with previous theoretical studies \citep{Kasen_2009,2004cetd.conf...43H}. Lower explosion energies produce less luminous light curves with longer plateau durations, while the luminosity of the radioactive tail is strongly governed by the synthesized $^{56}\mathrm{Ni}$ mass. These relationships provide useful constraints on the physical properties of the progenitor and explosion.

Despite these trends, significant degeneracies remain within the progenitor mass ($M$), radius ($R$), and explosion energy ($E$) parameter space. Future work combining photometric and spectroscopic modeling at multiple epochs may help break these degeneracies and place tighter constraints on the progenitor properties and explosion parameters of SN~2024abfl, possibly using additional constraints from the nebular phase. 

The unique features of this supernova call for more detailed theoretical studies of both photometric and spectroscopic observations. A lack of mixing could be an explanation for these features, while the explosion energy, progenitor radius and mass parameters also affect the extent of mixing within the ejecta. Further study is required to understand the physics needed to reproduce the results from this work, and to gain a better general understanding of what factors affect mixing of ejecta in Type II SNe.

\section{Acknowledgments} \label{sec:acknowledgments}

We would like to thank Madison Gerard, Azalee Bostroem, David Sand, and the rest of the Supernova Group at Arizona State Univeristy for providing us with the observational data for this work. Additionally, this publication was made possible due to external comments and feedback from Jared Goldberg of the Flatiron Institute, along with Tess Hoover and Colin Macrie of the Puffins Lab at Purdue University.

\pagebreak

\bibliography{bib}{}
\bibliographystyle{aasjournalv7}



\end{document}